\documentclass[10pt, conference, compsocconf]{IEEEtran}
\usepackage{color}
\usepackage{subfigure}
\usepackage{graphicx}
\usepackage{amssymb }
\usepackage{algorithm}
\usepackage{algpseudocode}
\usepackage{multirow}
\usepackage[numbers,sort&compress]{natbib}
\newenvironment{sequation}{\small\begin{equation}}{\end{equation}}

\usepackage{verbatim}

\ifCLASSINFOpdf
\else
\fi
\begin{document}
%
% paper title
% can use linebreaks \\ within to get better formatting as desired
%\title{$COL-KV^{\alpha}$: A Collaborative Key-Value Store Using Near-Data Processing to Improve Compaction }
%\title{COL-KV: Col-KV  ColKV CoKV }
%\title{Co-KV$_{s}$: A \textit{Co}llaborative \textit{K}ey-\textit{V}alue Store Using Near-Data Processing to Improve Compaction for the LSM-tree}
\title{Co-KV: A \textit{Co}llaborative \textit{K}ey-\textit{V}alue Store Using Near-Data Processing to Improve Compaction for the LSM-tree}
% author names and affiliations
% use a multiple column layout for up to two different
% affiliations

\author{\IEEEauthorblockN{Hui Sun, Wei Liu}
\IEEEauthorblockA{School of Computer Sci. and Tech.\\
Anhui University\\
Hefei China\\
sunhui@ahu.edu.cn, liuweiahu@gmail.com}
\and
\IEEEauthorblockN{Jianzhong Huang}
\IEEEauthorblockA{Wuhan National Lab. for Optoelectronics\\
 Huazhong University of Sci. and Tech.\\
 Wuhan China \\
 hjzh@hust.edu.cn
}
\and
\IEEEauthorblockN{Weisong Shi}
\IEEEauthorblockA{Department of Computer Science\\
 Wayne State University\\
 Detroit, MI, USA\\
 weisong@wayne.edu
}
}
%\vspace{-30pt}

% conference papers do not typically use \thanks and this command
% is locked out in conference mode. If really needed, such as for
% the acknowledgment of grants, issue a \IEEEoverridecommandlockouts
% after \documentclass

% for over three affiliations, or if they all won't fit within the width
% of the page, use this alternative format:
% 
%\author{\IEEEauthorblockN{Michael Shell\IEEEauthorrefmark{1},
%Homer Simpson\IEEEauthorrefmark{2},
%James Kirk\IEEEauthorrefmark{3}, 
%Montgomery Scott\IEEEauthorrefmark{3} and
%Eldon Tyrell\IEEEauthorrefmark{4}}
%\IEEEauthorblockA{\IEEEauthorrefmark{1}School of Electrical and Computer Engineering\\
%Georgia Institute of Technology,
%Atlanta, Georgia 30332--0250\\ Email: see http://www.michaelshell.org/contact.html}
%\IEEEauthorblockA{\IEEEauthorrefmark{2}Twentieth Century Fox, Springfield, USA\\
%Email: homer@thesimpsons.com}
%\IEEEauthorblockA{\IEEEauthorrefmark{3}Starfleet Academy, San Francisco, California 96678-2391\\
%Telephone: (800) 555--1212, Fax: (888) 555--1212}
%\IEEEauthorblockA{\IEEEauthorrefmark{4}Tyrell Inc., 123 Replicant Street, Los Angeles, California 90210--4321}}

% use for special paper notices
%\IEEEspecialpapernotice{(Invited Paper)}

% make the title area
\maketitle

\begin{abstract}
Log-structured merge tree (LSM-tree) based key-value stores are widely employed in large-scale storage systems. In the compaction of the key-value store, SSTables are merged with overlapping key ranges and sorted for data queries. This, however, incurs write amplification and thus degrades system performance, especially under update-intensive workloads. Current optimization focuses mostly on the reduction of the overload of compaction in the host, but rarely makes full use of computation in the device. To address these issues, we propose Co-KV, a \textit{Co}llaborative \textit{K}ey-\textit{V}alue store between the host and a near-data processing ({\em i.e.,} NDP) model based SSD to improve compaction. Co-KV offers three benefits: (1) reducing write amplification by a compaction offloading scheme between host and device; (2) relieving the overload of compaction in the host and leveraging computation in the SSD based on the NDP model; and (3) improving the performance of LSM-tree based key-value stores under update-intensive workloads.

Extensive db\_bench experiment show that Co-KV largely achieves a 2.0x overall throughput improvement and a write amplification reduction by up to 36.0\% over the state-of-the-art LevelDB. Under YCSB workloads, Co-KV increases the throughput by 1.7x $\sim$ 2.4x while decreases the write amplification and average latency by up to 30.0\% and 43.0\%, respectively. 
\end{abstract}


% For peer review papers, you can put extra information on the cover
% page as needed:
% \ifCLASSOPTIONpeerreview
% \begin{center} \bfseries EDICS Category: 3-BBND \end{center}
% \fi
%
% For peerreview papers, this IEEEtran command inserts a page break and
% creates the second title. It will be ignored for other modes.
\IEEEpeerreviewmaketitle

\section{Introduction}
\label{Introduction}
% no \IEEEPARstart
Applications with many types of data are exploding in the big data era. IDC \cite{turner2014digital} reveals that the digital universe is large and is doubling in size every two years. Data-intensive applications must be supported by high performance. In relational databases, retrieval \cite{han2011survey} is too inefficient to provide high performance for large-scale data processing. Key-value stores with high performance are widely employed in storage and memory systems \cite{borthakur2013under}. Many famous IT companies built their data infrastructure using LSM-tree based key-value store \cite{o1996log}, \textit{e.g.,} Google with LevelDB \cite{leveldb2014fast}.  

However, the compaction procedure, in which SSTables are merged with overlapping key ranges and ordered for data queries, is a weak link for LSM-tree based key-value stores. Compaction causes write amplification \cite{o1996log} and large amounts of data movement between host and device, which degrades performance, especially under update-intensive workloads. Current optimizations for LSM-tree based key-value stores concentrate mostly on two aspects. Firstly, the approaches on the host side aim to reduce I/O accesses to the device during compaction \cite{wu2015lsm}. The host-side parallelism of CPU and I/O resources is used to improve compaction \cite{zhang2014pipelined}. The CPU usage in the host increases while the computation resources in storage device are far from being fully utilized. Secondly, a prior method \cite{yangintorage} offloads the whole compaction from the host into the devices to optimize compaction. This approach avoids frequent data movement but relies on CPU computation in the resource-constrained devices, thus its inability to maximize the overall system performance.

The popular framework moving computing close to source data is named near-data processing ({\em i.e.,} NDP) in storage level \cite{Balasubramonian2016Near} or processing-in-memory in memory level \cite{khoram2017challenges}\cite{Ahn2015A}, respectively. This model aims to reduce the cost of data movement in the system under big-data workloads. In-storage computing model \cite{yangintorage} applied in NAND flash-based storage system is widely studied. With computing ability increasing, new frameworks \cite{do2013query}\cite{gu2016biscuit}\cite{kang2013enabling} with storage and computing pre-process the tasks collaboratively with the host, which {reduces data movement and thus relieves the load of bandwidth and improves} system performance. Our work aims to study the optimization of compaction in a key-value store using the storage-level NDP model. 

A \textit{Co}llaborative \textit{K}ey-\textit{V}alue store, referred to hereafter as Co-KV, is proposed for compaction optimization. 
%Besides, Co-KV supports the full benefits of LevelDB. 
Contributions of this paper include:
\textit{
(1) NDP model-based Co-KV can leverage computation resources in host and device and the parallelism in the overall system; 
(2) Co-KV enables signification improvement in compaction by means of a compaction-offloading collaborative scheme between host and device; and
(3) With Co-KV in place, overload of the host CPU and I/O resources is relieved and the overall performance is optimized in key-value stores.
%(1) Using NDP model, a novel key-value store ({\em i.e.,} Co-KV), collaborating computation resources in host and device, is proposed to leverage the computation and parallelism in the overall system;
%(2) Compaction is largely improved by means of a dynamic compaction offloading scheme between host and device in Co-KV; 
%(3) Overload of host CPU and I/O resources in key-value storage system is relieved and overall performance is optimized.
}

%The rest of this paper is organized as follows. Section \ref{BACKRE} presents background and related work for LSM-tree based key-value store and near-data processing model. Section \ref{COLKVdesg} and \ref{COLKVimp} give details of Co-KV. 
%Experiment and evaluation are discussed in Section \ref{evaluation}. Finally, we make a conclusion after discussion and future work about Co-KV.

\section{Background and Related Work}
\label{BACKRE}
This section presents the background and related work on LSM-tree based key-value stores and storage-level near-data processing, which are helpful for us to design Co-KV. 
\subsection{LSM-tree based Key-Value Stores}
LSM-tree based key-value stores \cite{borthakur2013under}\cite{leveldb2014fast} provide high throughput in large-scale storage systems, LevelDB, for example. A key-value item is written into a commit log and buffered in a MemTable in memory. When the size of MemTable reaches its maximum capacity, it is transformed into an immutable MemTable and dumped into a disk as the SSTable. SSTables are organized into a series of levels that grow exponentially in size. An SSTable is migrated from $L_{k}$ to $L_{k+1}$ level, a process termed \textbf{\textit{compaction}}. Compaction affects the performance of key-value store by incurring write amplification which is the ratio of the amount of data written to the underlying storage device to that requested by the from the user. Written data volume in $L_{k+1}$ level is about ten times as large as that in $L_{k}$ in the worst case during a compaction.

Most schemes have been proposed on the host side. bLSM \cite{sears2012blsm} with the benefits of the LSM- and B-tree has a merge scheduler to improve performance. VT-tree \cite{shetty2013building} eliminates unnecessary copying SSTables into new ones in compaction. RocksDB \cite{borthakur2013under} triggers more than two contiguous levels at once to sort and merge, which reduces compaction. LSM-trie \cite{wu2015lsm} decreases write amplification by improving the random performance on multiple sorted tables in each level. The cLSM \cite{Golan2015Scaling} and pipelined compaction procedure \cite{zhang2014pipelined} employs concurrent or parallelism to improve the compaction performance. WiscKey \cite{lu2016wisckey} integrates the internal parallelism of SSD with the key-value separation method to achieve a high query performance. Atlas \cite{lai2015atlas} is a distributed key-value store that stores keys and values on different HDDs. LSbM \cite{tenglsbm} introduces a buffer management in the disk to minimize cache invalidation caused by compaction in the device. 
 
In contract, our Co-KV makes full use of computation in host-side and device-side subsystems to substantially improve the overall system performance degraded by compaction. 
%%
%\begin{figure}[t]
%\setlength{\abovecaptionskip}{0.cm}
%\setlength{\belowcaptionskip}{-0.cm}
%\centering
%\includegraphics[height=1in, width=1.6in]{Fig1-leveldb}
%\caption{Architecture and Compaction process in LevelDB.}
%\label{Fig1}
%\vspace{-18pt}
%\end{figure}
\subsection{Near-Data Processing Model} 
%%%%%%%%%%%%%%%%%%%%%%%%%%%%%%%%%%%%%%
%%%%%%%%%%%%%%%%%%%%%%%%%%%%%%%%%%%%
In the era of big data, large-scale systems change from a computing-intensive to a data-intensive processing model. In the computing-intensive model, performance and energy \cite{wulf1995hitting} of the system are increasingly consumed by the high overload of data movement \cite{balasubramonian2014near}. The I/O interface between storage and memory is always the bottleneck in high-performance systems \cite{yangintorage}, which are evenly aggravated by data movement from storage to memory. To address these issues, a new notion of moving computation to data is proposed and named near-data processing ({\em i.e., NDP}) \cite{gu2016biscuit}\cite{liu2013computing}. 

{Our work focuses mainly on the storage-level NDP}. The active disk \cite{seminar1998active} is a prior prototype of NDP model in storage level. With the advent of the flash memory, the framework of the active disk attracts attention from both academia and industry for two reasons: (1) The bandwidth of the flash memory is much higher than that required by the I/O between host and device \cite{do2013query}; (2) Inside a flash memory-based device, the computation controller, which mainly runs flash translation layer, can be enhanced to deal with computing-intensive tasks \cite{cho2013xsd}\cite{Wang2016SSD}. The notion of storage-level NDP (or in-storage computing \cite{cho2015solid}\cite{Park2016In}) has been increasingly employed in the big-data workloads. 

Therefore the NDP model is applied into our novel Co-KV to improve the compaction of LSM-tree based key-value stores by host- and device-side collaboration. 
%\textit{\textbf{Memory-level NDP}}: Processing-in-memory (PIM)  in the 1990s failed to widely employed in industry because of its incompatible technologies for manufacturing DRAM and CPUs \cite{Balasubramonian2016Near}. With the emerge of 3D memory package technology \cite{Jeddeloh2012Hybrid}, one or more DRAM dies is integrated with low-power computational resources close to the memory, which regains attention on PIM from academic and industry. PIM  decreases data movement by placing computation close to memory in which data resides. This can relieve the challenging of memory wall  in computer architecture. 
%%%%%%%%%%%%%%%%%%%%%%%%%%%%%%%%%%%%%%%%%

\section{Co-KV Design}
\label{COLKVdesg}
%Write amplification due to compaction lowers overall system performance in LSM-tree key-value store, which motivated us to design and implement Co-KV store system.  
%The Co-KV is designed by extending a popular open source LevelDB library in this paper.
%%%%
%%%% 
\subsection{Motivation}
LSM-tree based key-value stores offer high performance and low latency. \textit{\textbf{Compaction}} is the bottleneck of performance because of movement of huge amounts of data between host and device. The storage-level NDP model moves computation to data and gives the chance to process data in a storage device. The whole or part of the task is handled in the store, with the intermediate or final result passed to the host. Data are processed on both sides in parallel, as a result of which, CPU consumption and I/O resources of the host are relieved. But, how to divide compaction tasks into host and storage device? How to synchronize both-side tasks and manage data consistency? These challenges motivate us to design a collaborative compaction for LSM-tree based key-value store using the benefits of the NDP model.

Then, we propose a new key-value store prototype, called Co-KV, by collaboration of computation in host and device to perform compaction in parallel. It significantly improves the performance of the key-value store. 

\subsection{Design Goals}
Co-KV employs the NDP model to handle the challenges of compaction in LSM-tree based key-value stores. Co-KV is designed with the following goals in mind:
%
%(1) How to implement the method of dynamic compaction offloading in both sides to effectively improve the ability of collaboration in the Co-KV store? 
%
%(2) How to design the collaborative scheme between the host and the device to utilize their computation resources to perform compaction in parallel in the Co-KV store?

\textbf{Low write amplification.} Write amplification induces extra data movement from device to host during compaction. It largely impacts overall performance; thus, it is essential to minimize write amplification for better performance of the key-value store.

\textbf{Reduction of CPU and I/O overload in host.} Data moving from device for compaction consumes CPU and I/O resources in the host. Co-KV aims to reduce the burden of CPU and I/O resources on the host side by offloading compaction into the device to enhance the overall performance.

\textbf{Optimized host- and device-side computation.} Compaction costs computation in the host, while the computation in devices is rarely exploited. Based on the NDP model, Co-KV executes compaction between host and device using the computation resources of both for high performance.

\subsection{Prototype of the Co-KV System}
\label{COKVPRO}
We introduce the framework and implementation of Co-KV (see Figure \ref{Fig2}) with emphasis on two key technologies: compaction offloading and collaborative management. 

\subsubsection{\textbf{Overall System of Co-KV}}
\begin{figure}[!t]
\setlength{\abovecaptionskip}{0.cm}
\setlength{\belowcaptionskip}{-0.cm}
\centering
\includegraphics[height=1.2in, width=2.4in]{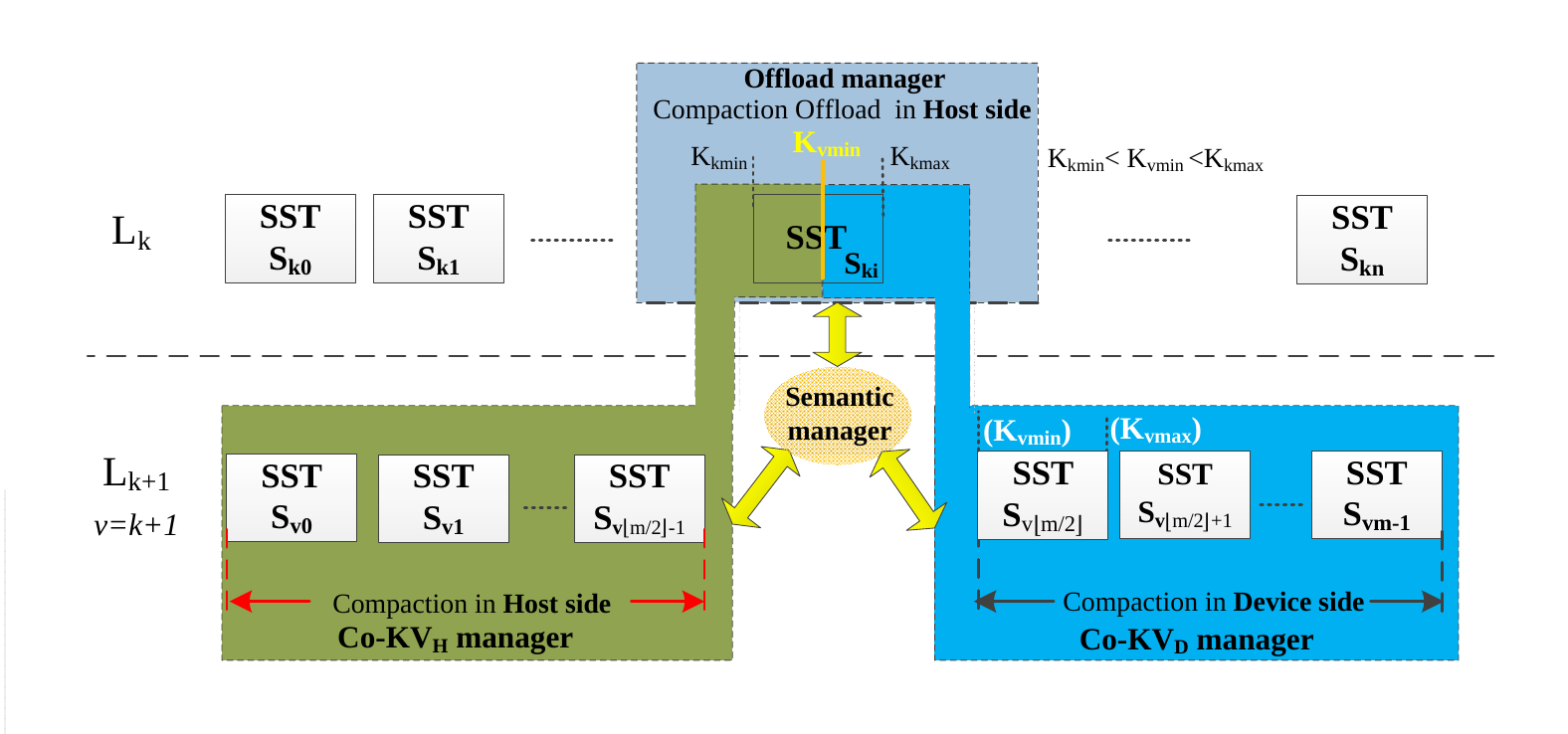}
\caption{Overall System of Co-KV.}
\label{Fig2}
\vspace{-18pt}
\end{figure}
%9-6
We present the main components related to the collaborative compaction in $L_{k}$ and $L_{k+1}$ levels between host and device. Co-KV is mainly composed of the host-side compaction manager (\textit{i.e.,} \textit{Co-K}$V_H$ manager), the device-side compaction manager (\textit{Co-K}$V_D$ manager for short), the compaction offloading manager (i.e., Offload manager), and the Semantic manager. 

{Offload Manager}: The Offload manager is an essential element in Co-KV. A scheme of offloading compaction named \underline{C}ompaction-\underline{S}STables-\underline{A}ware (CSA for short) is implemented. The Offload manager splits meta-data of compaction into host- and device-side ones. Meta-data, as the semantic format, is transmitted to the host- and device-side managers respectively through the semantic manager. Meanwhile, the Offload manager receives the meta-data from the two compaction managers and integrates them into a new one, which maintains meta-data consistency during the compaction. The Offload manager deals with the compaction overload between the host-side and device-side systems, and therefore reduces write amplification and enhances the system performance.

{\textit{Co-K}$V_H$ Manager}: A host-side compaction manager in Co-KV ({\em i.e.,} \textit{Co-K}$V_H$ in Figure \ref{Fig2}) is fully compatible with application interfaces (put, get, delete, etc.) in LevelDB. The \textit{Co-K}$V_H$ manager is responsible for host-side compaction, data consistency, and meta-data management. Meta-data for compaction is offloaded from the Offload manager (see Offload in Figure \ref{Fig2}) to the \textit{Co-K}$V_H$ manager, which performs host-side compaction operations. Performance in the host can be improved by reducing the scale of compaction operations with respect to that in LevelDB. Results of compaction ({\em e.g.,} FileMetaData) are returned to the Offload manager via semantic messages by a semantic manager.

{\textit{Co-K}$V_D$ Manager}: The device-side compaction manager (\textit{Co-K}$V_D$) exemplifies the near-data processing model in the device level. The \textit{Co-K}$V_D$ manager receives a portion of the meta-data in a semantic message from the Offload manager via the semantic manager. In this way, the \textit{Co-K}$V_D$ manager locally performs compaction using the NDP model rather than sends more SSTables to the host. A portion of compactions in the device contributes to low write amplification in Co-KV and high overall system performance. File meta-data (key range, file number, and file size) of new SSTables produced from device-side compaction is returned to the host. 

{Semantic Manager}: The Semantic manager is an important part of Co-KV for the collaborative compaction between the host and the device, {\em i.e.,} the collaborative compaction in the \textit{Co-K}$V_H$ manager and the \textit{Co-K}$V_D$ manager. The Semantic message is the basic data transmission unit of meta-data for compaction between the Offload manager and the \textit{Co-K}$V_H$ manager (or the \textit{Co-K}$V_D$ manager). The semantic manager plays an important role in keeping the data consistency and optimizing parallel compaction operations between the above two compaction managers. 

Details of the implementation for these components in Co-KV are demonstrated in the following section.

\section{{Co-KV Implementation}}
\label{COLKVimp}
We introduce the scheme of compaction offloading for the \textit{Co-K}$V_H$ manager and the \textit{Co-K}$V_D$ manager. The collaborative compaction between these two managers is described.

In LevelDB, an SSTable ({\em e.g.,} $S_{ki}$ SSTable) in $L_{k}$ level and a number of SSTables in $L_{k+1}$ level are read from the device when a compaction is triggered. Meta-data are updated after SSTables being written back to the device. In comparison, Co-KV offloads a portion of compaction onto the device. The NDP model is employed by the \textit{Co-K}$V_D$ manager in the device to perform compaction by using its computation and a lightweight run-time library . 

It should be noted how the collaborative compaction managers between two sides handle SSTables during compaction. Two challenges must be handled: \\
\textit{\noindent\ $\diamond$ How can the Offload manager select SSTables as action objects during compaction? ({\textit{\textbf{Problem 1}}}) \\
\noindent\ $\diamond$ How can the Offload manager split and offload SSTables to \textit{Co-K}$V_H$ and \textit{Co-K}$V_D$ managers? ({\textit{\textbf{Problem 2}}})}
%%%%
%In comparison, Co-KV store dynamically offloads portion of compaction to the device and performs compaction in parallel in both host- and device-side subsystem. Co-KV store leverages computation in the device and parallelism between the host and the device to relieve the consumption of computation and I/O resources in the host during the processing of compaction. Overload of compaction is lowered and performance of key-value store is largely improved. 

\textbf{$\triangleright$Problem 1.} Experiments are conducted to analyze the characteristics of compaction in LevelDB and find how many SSTables should be selected from $L_{k}$ and $L_{k+1}$ levels by the Offload manager. Since compaction in LevelDB results mainly from the random workload, db\_bench in LevelDB is employed in this test. Under fillrandom workload, a data set with one billion key-value items is configured to test the number of SSTables from $L_{k}$ level in compaction.

\textbf{Observation:}
We find in most cases that more than 93.0\% single SSTables in $L_{k}$ level and no less than 80\% double SSTables in $L_{k+1}$ level are selected to be compacted. These observations lead us to (1) select one SSTable from $L_{k}$ as the number of SSTables in our Co-KV; and (2) design an approach to offloading half of the SSTables in $L_{k+1}$ to the device-side compaction manager by the Offload manager. The total of compaction SSTables is reduced on the host side. Meanwhile, host- and device-side compactions are executed in parallel, which greatly improves the performance. This approach makes full use of computation in the overall system and proves to be efficient in this section and in Section \ref{evaluation}. This approach is used in most cases to improve compaction for the overall system, with the original method employed for the rest ones. 
%More detail about performance improvement is shown in Section \ref{EVALU}.
%%%%%%%%%%%%

\textbf{$\triangleright$Problem 2.} We design a scheme of offloading compaction in Offload manager to (1) improve compaction and lower write amplification; and (2) employ the NDP model in the device to achieve maximum parallelism in compaction. This scheme refered to as \textit{C}ompaction-\textit{S}STables-\textit{A}ware (\textit{\textbf{CSA}}) offloads compaction to the host- and device-side compaction managers. The Offload manager obtains meta-data of SSTables across its levels and SSTables in the current version for compaction. The CSA scheme is illustrated as follows.  

%\footnotesize
It is supposed that $S_{ki}$ SSTable in $L_{k}$ level is involved in compaction. The key range of $S_{ki}$ SSTable is given by
\begin{sequation}
\setlength{\abovedisplayskip}{2pt}
\setlength{\belowdisplayskip}{2pt}
\label{E1}
R_{S_{ki}} = \left[K_{kmin}, K_{kmax}\right],
\end{sequation}
%%%%%%%%%%%%%%%%%%
which is applied to compare with the key range of every SSTable in $L_{k+1}$ level to find the SSTable with overlapping one by
 %%%%%%%%%%%%%%%%%
\begin{sequation}
\setlength{\abovedisplayskip}{2pt}
\setlength{\belowdisplayskip}{2pt}
\label{E2}
R_{S_{ki}} \cap R_{S_{vj}} \neq \emptyset , (v = k+1).
\end{sequation}
%%%%%%%%%%%%%%%%
where $0\leq i \leq n$ and $0\leq j \leq m-1$ as shown in Figure\ref{Fig2}.

Then, the key-range set of SSTables that have an overlapping key range with $S_{ki}$ SSTable is involved in the process of compaction. These ranges of key in $L_{k+1}$ level can be expressed as
%%%%%%%%%%%%%%
\begin{sequation}
\setlength{\abovedisplayskip}{2pt}
\setlength{\belowdisplayskip}{2pt}
\label{E3}
Set_{comp} = \left\{R_{S_{v0}}, R_{S_{v1}}, \cdots, R_{S_{vm-1}}\right\},
\end{sequation}
where \textit{v = k+1}.

By the CSA scheme, elements in $Set_{comp}$ are divided into two subsets: 
%%%%%
\begin{sequation}
\setlength{\abovedisplayskip}{2pt}
\setlength{\belowdisplayskip}{2pt}
\label{E4}
Set_{comp_{sub1}} = \left\{R_{S_{v0}}, R_{S_{v1}}, \cdots, R_{S_{v\left\lfloor m/2\right\rfloor-1}}\right\}, 
\end{sequation}
%%%%%%%%%%%
%%
where there is \textit{v = k+1} and 
\begin{sequation}
\setlength{\abovedisplayskip}{2pt}
\setlength{\belowdisplayskip}{2pt}
\label{E5}
Set_{comp_{sub2}} = \left\{R_{S_{v\left\lfloor m/2\right\rfloor}}, R_{S_{v\left\lfloor m/2\right\rfloor}+1}, \cdots, R_{S_{vm-1}}\right\},
\end{sequation}
%%%%%%%%%%%
where \textit{v = k+1} and {$R_{S_{v\left\lfloor m/2\right\rfloor}} = \left[K_{vmin}, K_{vmax}\right]$}.

%%%%%%%%%%%%%%
Then key range $R_{S_{ki}}$ for $S_{ki}$ SSTable is split by the value of the lower range of $R_{S_{v\left\lfloor m/2\right\rfloor}}$ ({\em i.e.,} $k_{vmin}$), and $R_{S_{ki}}$ can be expressed as
\begin{sequation}
\setlength{\abovedisplayskip}{2pt}
\setlength{\belowdisplayskip}{2pt}
\label{E7}
R_{S_{ki}} = \left[K_{kmin}, K_{vmin}\right] \cup \left[K_{vmin}, K_{kmax}\right],
\end{sequation}
where $R_{S_{ki}}$ is rewritten into 
\begin{sequation}
\setlength{\abovedisplayskip}{2pt}
\setlength{\belowdisplayskip}{2pt}
\label{E8}
R_{S_{ki}} = R_{S_{ki}-t} \cup R_{S_{ki}-b},
\end{sequation}
where {$R_{S_{ki}-t} = \left[K_{kmin}, K_{vmin}\right]$ and $R_{S_{ki}-b} = \left[K_{vmin}, K_{kmax}\right]$}.

Finally, two sets of SSTables are obtained by the Offload manager using a CSA scheme: 
\begin{sequation}
\setlength{\abovedisplayskip}{2pt}
\setlength{\belowdisplayskip}{2pt}
\label{E9}
Set_{comp-h} = \left\{Set_{comp_{sub1}},  R_{S_{ki}-t} \right\},
\end{sequation}
%%%%%%%
and 
%%%%%%%
\begin{sequation}
\setlength{\abovedisplayskip}{2pt}
\setlength{\belowdisplayskip}{2pt}
\label{E10}
Set_{comp-d} = \left\{Set_{comp_{sub2}}, R_{S_{ki}-b} \right\}.
\end{sequation}
When compaction is offloaded into the two managers, the collaborative compaction is performed on both sides. 
A \textbf{\textit{semantic manager}} is designed to handle the semantic transmission between the host- and device-side compaction managers. When the meta-data of SSTables in compaction is dispatched to the device, the semantic manager is triggered to send the meta-data to the device. In compaction, the \textit{Co-K}$V_H$ manager (or \textit{Co-K}$V_D$ manager) receives and checks the semantic requests or responses from its counterpart. The semantic manager goes inactive upon the completion of the semantic transmission from both sides. An example is given to illustrate the semantic management in compaction. 

When the device-side compaction completes, the \textit{Co-K}$V_D$ manager generates a semantic with the completion flag and send it to \textit{Co-K}$V_H$. The meta-data after compaction is sent to the Offload manager by the semantic manager. When \textit{Co-K}$V_H$ compaction is done, \textit{Co-K}$V_H$ sends its new meta-data to the Offload manager and checks the completion flag in an incoming semantic from the device. If compaction has not been completed, the \textit{Co-K}$V_H$ manager continues to check the completion flag of semantic from the device until the compaction in device finishes. 
Owing to the semantic manager, the semantics from both compaction manager can cooperate with each other in real time. The parallelism of collaborative compaction between host and device can keep data consistent. It is stated that the semantic manager a lightweight management software, which has little impacts on system performance.

Both the \textit{Compaction-SSTables-Aware scheme} and the semantic manager play an important role in our Co-KV. We put forward the procedures for \textit{Co-K}$V_H$ and \textit{Co-K}$V_D$ (see Algorithm \ref{alg1} and \ref{alg2}) combining two aspects.\\
%%%%%%%%%%%%%%%
%++++++++++++++++algorithm
%+++++++++++++++
\begin{footnotesize}
\begin{algorithm}\footnotesize
\setlength{\belowcaptionskip}{-0.6cm}
  \begin{algorithmic}[1]
	\caption{Compaction in host-side Co-KV}\label{alg1}
    \Procedure{\textit{Co-K}$V_H$ Compaction}{}
		\If {$CompactionTrigger$ == TRUE} \\
		\quad  Select $S_{ki}$ SSTable in $L_{k}$\\
		\quad  keyRange = $\left[K_{kmin}, K_{kmax}\right]$\\
		\quad  m = Numbers of SSTables in this compaction in $L_{k+1}$\\
		\quad  Set $R_{S_{ki}-t} = \left[K_{kmin}, K_{vmin}\right]$, $R_{S_{ki}-b} = \left[K_{vmin}, K_{kmax}\right]$\\
		\quad  Set $Set_{comp_{sub1}} = \left\{R_{S_{v0}}, R_{S_{v1}}, \cdots, R_{S_{v\left\lfloor m/2\right\rfloor-1}}\right\}$\\
		\quad  Set $Set_{comp_{sub2}} = \small{\left\{R_{S_{v\left\lfloor m/2\right\rfloor}}, R_{S_{v\left\lfloor m/2\right\rfloor}+1}, \cdots, R_{S_{vm-1}}\right\}}$\\
		\quad  OffloadManager(ToCo-KV\_H, $R_{S_{ki}-t}$, $Set_{comp_{sub1}}$)\\
	  \quad  OffloadManager(ToCo-KV\_D, $R_{S_{ki}-b}$, $Set_{comp_{sub2}}$)	\\	
		\quad  SemanticManager(SEND, \textit{Co-K}$V_H$, CompMetadata\_H)\\
		\quad  SemanticManager(SEND, \textit{Co-K}$V_D$, CompMetadata\_D)\\
	  \quad  Co-KV\_HmanagerDoCompaction()\\
		\quad  SemanticManager(SEND, Offload, \textit{Co-K}$V_H$newMetadata)
	  %\quad  HCompactionFinish = TRUE
	\While {DCompFinish == FALSE}\\ 
    \quad  \quad  \quad \quad  \quad   continue
  \EndWhile\\
    \quad  DCompFinish = FALSE\\
		\quad  SemanticManager(RECV, Offload, \textit{Co-K}$V_H$newMetadata)\\
		\quad  SemanticManager(RECV, Offload, \textit{Co-K}$V_D$newMetadata)\\
	  \quad  Update meta-data and delete obsoleteFile by Offload manager
		\EndIf   
	  \EndProcedure
  \end{algorithmic}
	\vspace{-4pt}
\end{algorithm}
	\vspace{-20pt}
\end{footnotesize}
%%%%%%%%%%%%%%%%%
\begin{footnotesize}
\begin{algorithm}\footnotesize
\setlength{\belowcaptionskip}{-0.6cm}
  \begin{algorithmic}[1]
	\caption{Compaction in device-side Co-KV}\label{alg2}
    \Procedure{\textit{Co-K}$V_D$ Compaction}{}\\
		\quad SemanticManager(RECV, \textit{Co-K}$V_D$, CompMetadata\_D)\\
	  \quad Co-KV\_DmanagerDoCompaction()\\
	  \quad DCompFinish = TRUE\\
		\quad SemanticManager(SEND, Offload, \textit{Co-K}$V_D$newMetadata)
	  \EndProcedure
  \end{algorithmic}
	\vspace{-4pt}
\end{algorithm}
\end{footnotesize}

%\subsubsection{Improving }
%compaction

%%%
%
%  增加各种测试方法来凸显我们方案的优越性
%
\section{Experiment and Evaluation}
\label{evaluation}
Experiments are conducted to evaluate the benefits of Co-KV under workloads: (1) the collaborative feasibility of Co-KV and availability of the offloading compaction between host and device; and (2) the superiority of Co-KV in compaction improvement under workloads.

\subsection{Experimental Settings}
\textbf{\textit{Hardware:}} We designed an experiment platform to evaluate Co-KV. Experiment is conducted on a system with four Intel(R) Core(TM) CPU i5-6500@3.20GHz processors and 16GB memory. An Intel 535 SSD is employed as the storage device featured with 180GB capacity, 540MB/s Seq. Read, and 490MB/s Seq. Write.

\textbf{\textit{Software:}} The operating system is Ubuntu 16.04 LTS with ext4 as the default file system. LevelDB a popular LSM-tree key-value store is employed as the baseline in experiment. 
The Co-KV framework extended from LevelDB consists of two components: the host- and device-side software frameworks. On the host side, the software framework is derived mainly from LevelDB. The APIs in Co-KV are compatible with those in LevelDB. The Runtime library inside LevelDB is applied in the \textit{\textit{Co-K}$V_H$ manager}.

%As real NVM DIMMs are , we
%emulate NVM using the DRAM similar to prior works
%[49, 52, 6]. The access latency of DRAM is about 60 ns
%[49], and the write latency of the latest 3D-XPoint is ten
%times of DRAM [3]. Thus, we set the NVM write latency
%to 600 ns. We add extra write latency only once
%for each persist operation as described in Section 3.4.2.
%We add the long write latency of NVM using the x86
%RDTSCP instruction. We use the instruction to read the
%processor timestamp counter and spin until the counter
%reaches the configured latency. We do not add extra
%read latency for NVM as it has similar read latency with
%DRAM [28, 31]. The impact of longer NVM read latency
%is evaluated in Subsection 4.6.

\textbf{\textit{NDPsim emulation:}} In our work, a real-world device-side SSD with a near-data processing model inside is not available for us yet. We implement a function-level simulator for the near-data processing model (named \textit{\textbf{NDPsim} with 3,000 lines of code}) which is emulated by a process on the host side by a software framework. In Col-KV, NDPsim should have two features: (1) the time consumption\footnote{In a real-world NDP-based SSD, the time consumption is largely dependent on the number of SSTables from flash memory to DRAM.} from a device to NDPsim is subtracted to simulate the internal high throughput in compaction inside a NDP-based storage device; (2) NDPsim is configured with heavy compaction load to simulate the case where the computation ability inside NDP-based SSD is smaller than that in the host. NDPsim provides a runtime system for SSTables compaction by using a library similar to that in LevelDB. 

In Co-KV, these host- and device-side frameworks execute their tasks in two different processes. As shown in Section \ref{COKVPRO}, \textit{Co-K}$V_H$ and \textit{Co-K}$V_D$ managers perform compaction, separately. The \textit{Co-K}$V_H$ manager in a process perform compaction according to meta-data $Set_{comp-h}$ (see Eq.\ref{E9}) on the host side. Meanwhile, the \textit{Co-K}$V_D$ manager inside NDPsim runs compaction by $Set_{comp-d}$ (see Eq.\ref{E10}) in the other process to simulate the real-world NDP runtime environment by a software framework. Therefore, there is no interference between the compactions in the host-side framework and device-side NDPsim. NDPsim invokes the \textit{Co-K}$V_D$ manager to handle SSTables compaction on receiving mete-data sets $Set_{comp-d}$ (see Eq.\ref{E10}) for compaction from Offload manager. NDPsim performance can be influenced by the meta-data transmission between the \textit{Co-K}$V_H$ manager in the host-side framework and its \textit{Co-K}$V_D$ manager, which is mainly attributable to inter-process communication. The \textit{{Offload manager}} collects meta-data for new SSTables on the completion of compaction by the two managers. Results from the \textit{Co-K}$V_H$ and \textit{Co-K}$V_D$ managers are written to different files in the device. The computation in the devices will not incur overloads on throughput of concurrent accessing the device. Finally, the meta-data from two compaction managers are written to a log file in the device by the Offload manager. 

Our goal is to show that the NDP model inside an SSD can improve compaction better in Co-KV than in LevelDB. NDPsim indicates the benefits of Co-KV with the near-data processing model. Work on the real-world NDP framework is under way to add computing components into SSD-based storage for compaction improvement in key-value stores.
%, and implementation of the programming model and runtime system for the NDP-based framework are in progress, as well. 

%We have dealt with the latency caused by the simulator to make results closest to that in the real-world environment.

\begin{figure*}[!t]\footnotesize
\setlength{\abovecaptionskip}{0.cm}
\setlength{\belowcaptionskip}{-0.cm}
\centering
\subfigure[]{
\includegraphics[height=1.1in, width=1.55in]{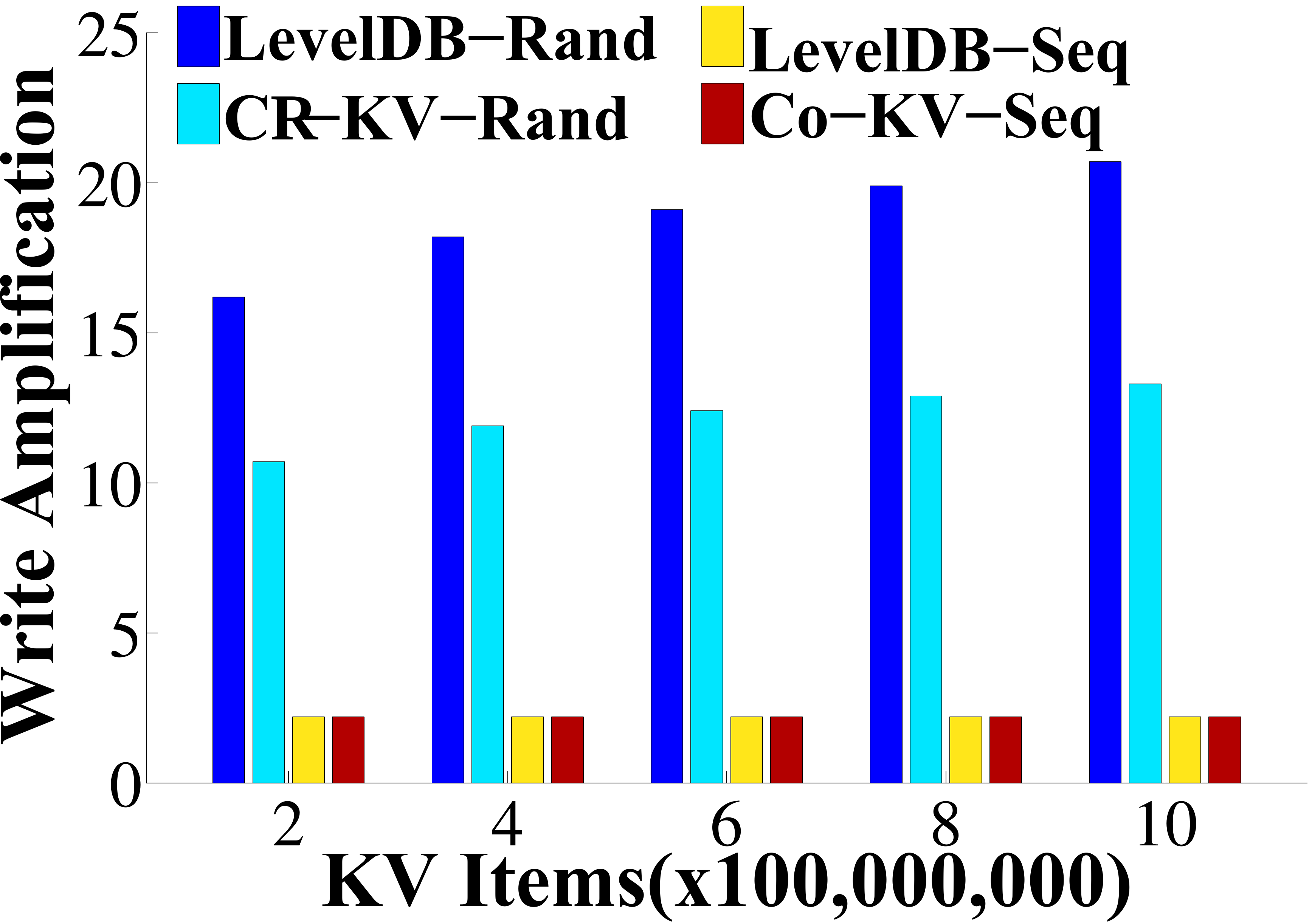}
\label{seq-random-WAR}
}
\subfigure[]{
\includegraphics[height=1.1in, width=1.55in]{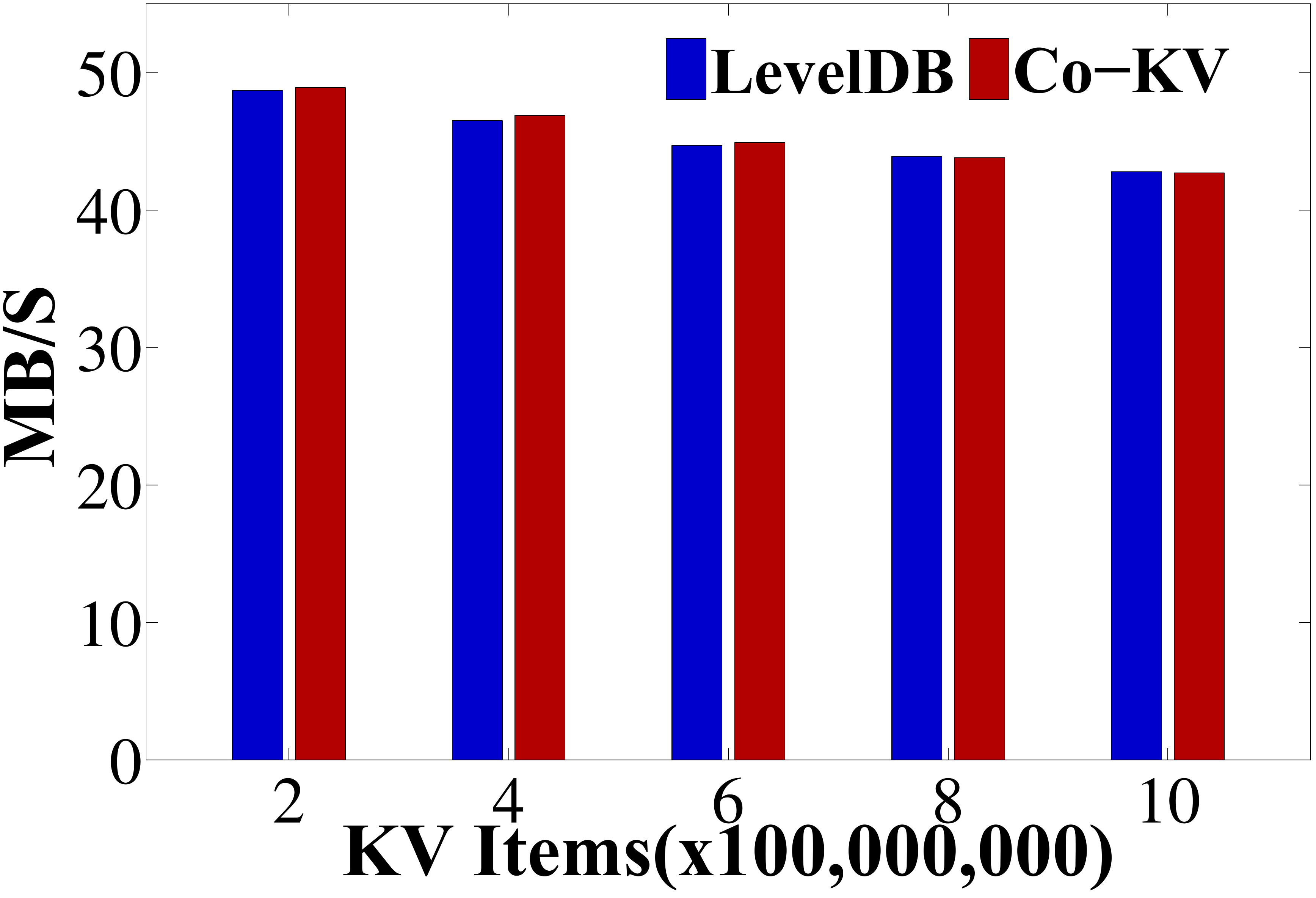}
\label{fillseqOMB}
}
\subfigure[]{
\includegraphics[height=1.1in, width=1.55in]{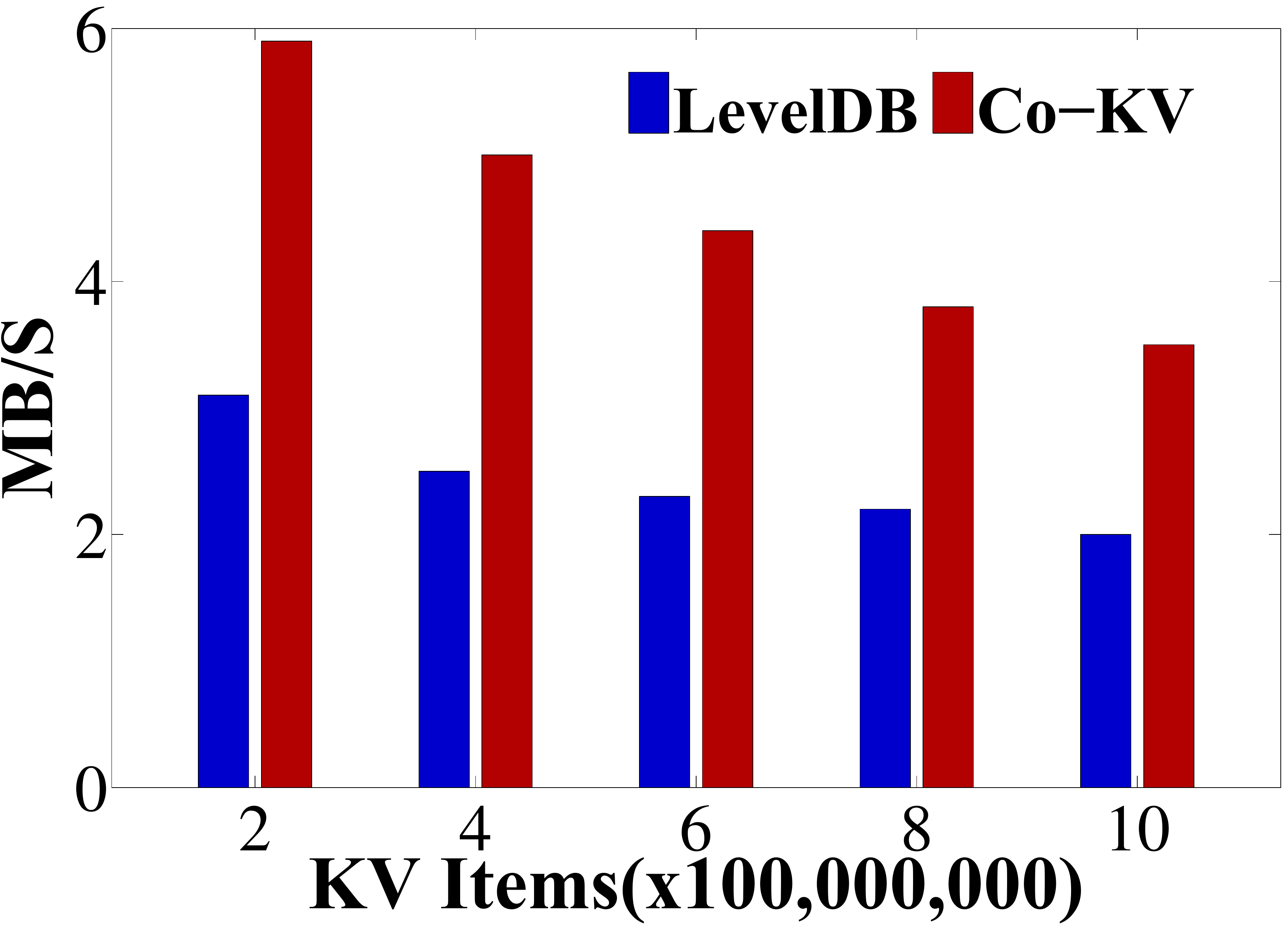}
\label{fillrandoMB}
}
%\subfigure[]{
%\includegraphics[height=1.2in, width=1.70in]{fillseqOCPU}
%\label{seq-random-cpu}
%}
\caption{Write Amplification (\ref{seq-random-WAR}), Throughput (MB/s, \ref{fillseqOMB} and \ref{fillrandoMB})) for LevelDB and Co-KV. \ref{seq-random-WAR} under fillseq and fillrandom workloads, \ref{fillseqOMB} fillseq workload, \ref{fillrandoMB} fillrandom workload.}
\vspace{-14pt}
\end{figure*}

\subsection{Workloads}
%\begin{table}[!t]\footnotesize
%\setlength{\abovecaptionskip}{0.cm}
%\setlength{\belowcaptionskip}{-0.cm}
%%% increase table row spacing, adjust to taste
%\renewcommand{\arraystretch}{1.3}
%% if using array.sty, it might be a good idea to tweak the value of
%% \extrarowheight as needed to properly center the text within the cells
%\caption{Parameters in Co-KV and LevelDB}
%\label{Param}
%\centering
%% Some packages, such as MDW tools, offer better commands for making tables
%% than the plain LaTeX2e tabular which is used here.
%\begin{tabular}{|c|c|c|c|c|c|}
%\hline
%\textbf{Type} & MemTable & SSTable & Data Block & Key & Value\\
%\hline
%\textbf{Size} & 4 MB & 2 MB & 4KB & 16B & 100B \\
%\hline
%\end{tabular}
%\vspace{-5pt}
%\end{table}

Db\_bench in LevelDB test-suite is used to test Co-KV. We compare Co-KV with LevelDB in terms of write amplification and throughput. The size of MemTable, SSTable, and data block is set to 4MB, 2MB and 4KB, respectively. The key-value item size is set to 16bytes and 100bytes which is the default configuration of LevelDB. Two workloads with ten types of key-value items (see Table II) are employed in evaluation.

\begin{table}[t]\footnotesize
\setlength{\abovecaptionskip}{0.cm}
\setlength{\belowcaptionskip}{-0.cm}
%% increase table row spacing, adjust to taste
\renewcommand{\arraystretch}{1.3}
% if using array.sty, it might be a good idea to tweak the value of
% \extrarowheight as needed to properly center the text within the cells
\caption{Workload Characteristics}
\label{workload}
\centering
%% Some packages, such as MDW tools, offer better commands for making tables
%% than the plain LaTeX2e tabular which is used here.
\begin{tabular}{|c|c|c|c|}
\hline
\multicolumn{4}{|l|}{\textbf{db\_bench - Workload in LevelDB}} \\
\hline
Type&fillrandom & fillseq & Key-Value Items\\
& & &($10^{9}$)\\ 
\hline
%db\_bench\_1 & $ \checkmark  $ & & \multirow{2}*{2x, 4x, 6x, 8x, 10x}\\
db\_bench\_1 & $ \checkmark  $ & & 2x, 4x, 6x, \\

%\hline
\cline{1-3}
db\_bench\_2 & & $ \checkmark $ & 8x, 10x \\
\hline
\hline
\multicolumn{4}{|l|}{\textbf{User-defined Workload in YCSB}} \\
\hline
OPS (Read\%)& \multicolumn{2}{|c|}{Database Size (10GB)} & Value Size (1KB)\\
\cline{2-4}
  (Update : Read) & Load (1x)&  Run (2x) &   1x  \\   
\hline
%{\multirow{3}*{0.25x,1x, 16x, 64x, 128x}} {\multirow{3}*{1x, 2x, 3x, 4x, 5x}} 
%DF-A\footnote{User-defined workload in YCSB} & Update:100\% & \multirow{10}*{1x, 2x, 3x, 4x, 5x} & \multirow{10}*{0.25x, 1x, 16x, 64x, 128x} \\
%&Read: 0\%&&\\
%\cline{1-2}
%DF-B& Update: 100\%&&\\
    %& Read: 0\% &&\\
%\cline{1-2}
%DF-C& Update: 100\%&&\\
    %& Read: 0\% &&\\
%\cline{1-2}
%DF-D& Update: 100\%&&\\
    %& Read: 0\% &&\\
%\cline{1-2}
%DF-E& Update: 100\%&&\\
    %& Read: 0\% &&\\
%\hline
%\end{tabular}
%\end{table}
\hline
 10\%, 30\%, & 1x, 2x & 1x, 2x, & \\
   50\%, 70\%, 90\% &  & 3x, 4x, 5x & -\\
\hline
\end{tabular}
\vspace{-15pt}
\end{table}
%%%%%%%%%%%%%%%%%%%%%%%%%%%%%%%%%%%%%%%%%%%%%
YCSB (\textbf{Y}ahoo! \textbf{C}loud \textbf{S}erving \textbf{B}enchmark) \cite{cooper2010benchmarking} is designed to evaluate performance of NoSQL databases. There are six default kinds of workloads. The impact of compaction on  performance results mostly from random updates. Thus, the advantage of Co-KV is obvious where compaction is performed under update-intensive workloads. Five types of workload executors by extending workload class of YCSB are modified to simulate different update-intensive workloads. Configurations of the random key and value size, which have great impact on compaction, are employed in the experiment. We set the request distribution to Zipf (see Table II). In our experiment, YCSB is used to evaluate the operations per second (ops/sec), latency, and write amplification of Co-KV.

%%%%%%%%%%%%%%%%%%%%%%%
%%%%%
%%%%%
\subsection{Evaluation}
\label{EVALU}

\subsubsection{{\textbf{Experiment based on db\_bench}}}
\begin{figure*}[!t]\footnotesize
\setlength{\abovecaptionskip}{0.cm}
\setlength{\belowcaptionskip}{-0.cm}
\centering
\subfigure[]{
\includegraphics[height=1.1in, width=1.57in]{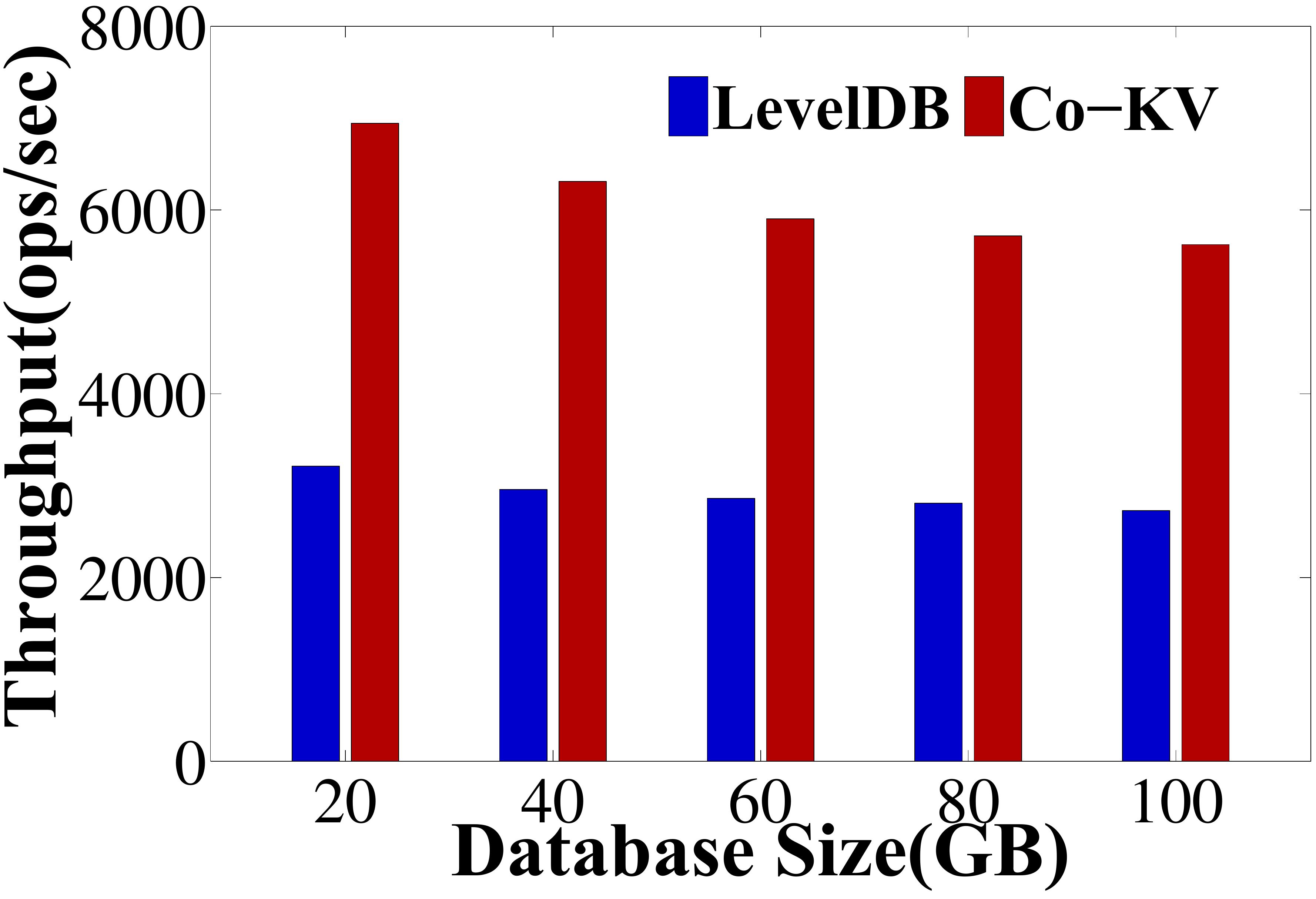}
\label{OPS-y1}
}
\subfigure[]{
\includegraphics[height=1.1in, width=1.57in]{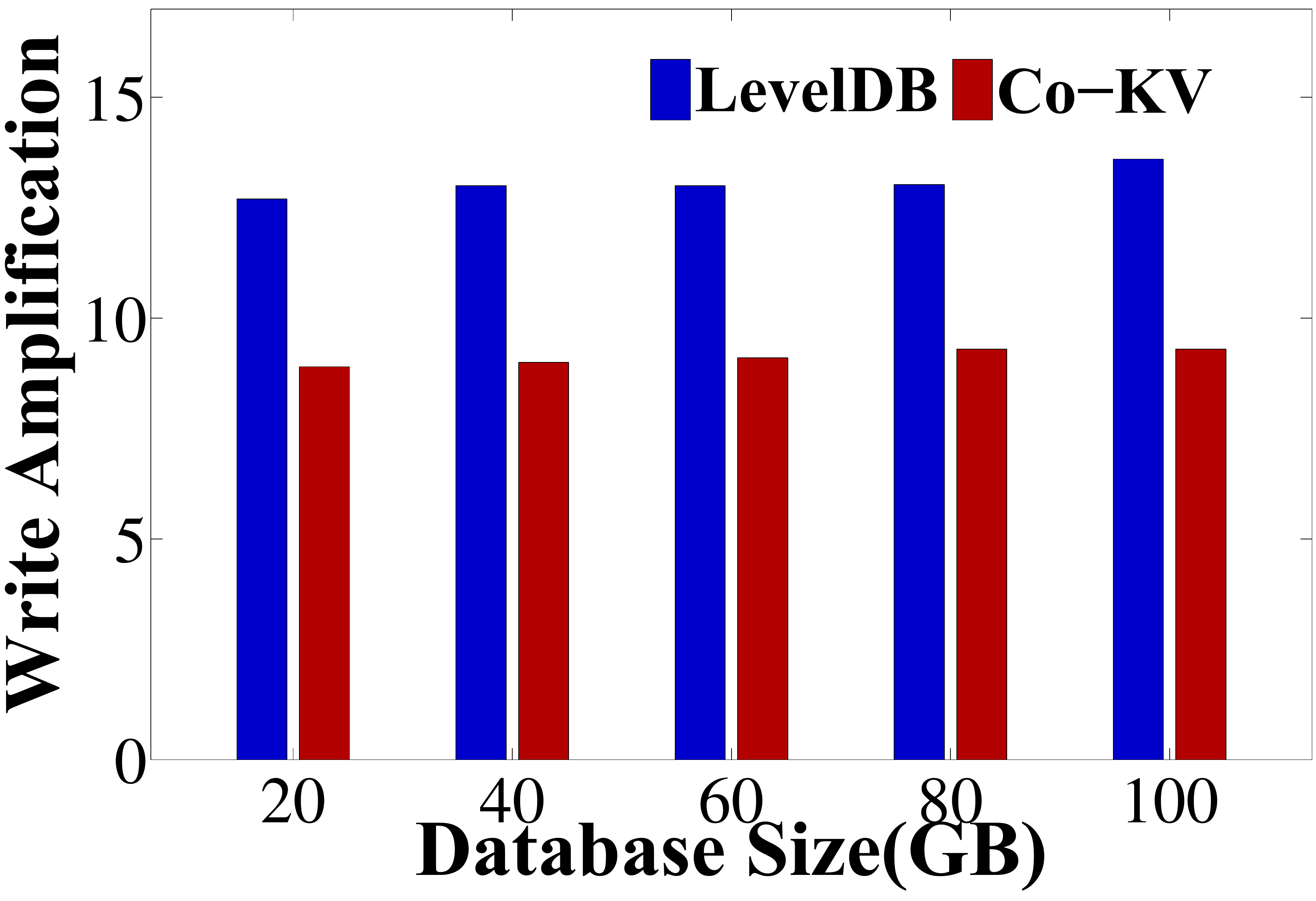}
\label{WAR-y1}
}
\subfigure[]{
\includegraphics[height=1.1in, width=1.57in]{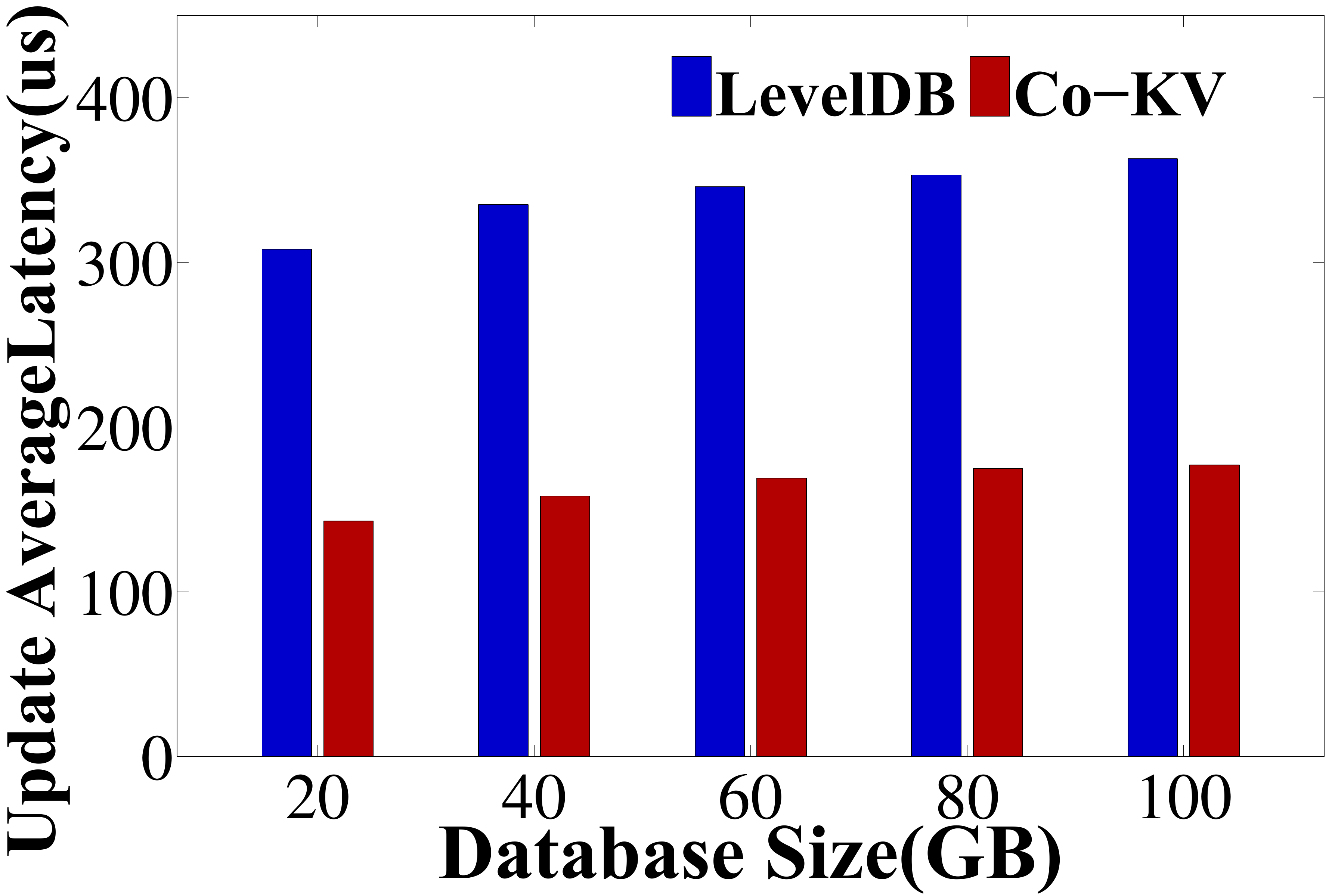}
\label{averageLatency-y1}
}
\subfigure[]{
\includegraphics[height=1.24in, width=1.57in]{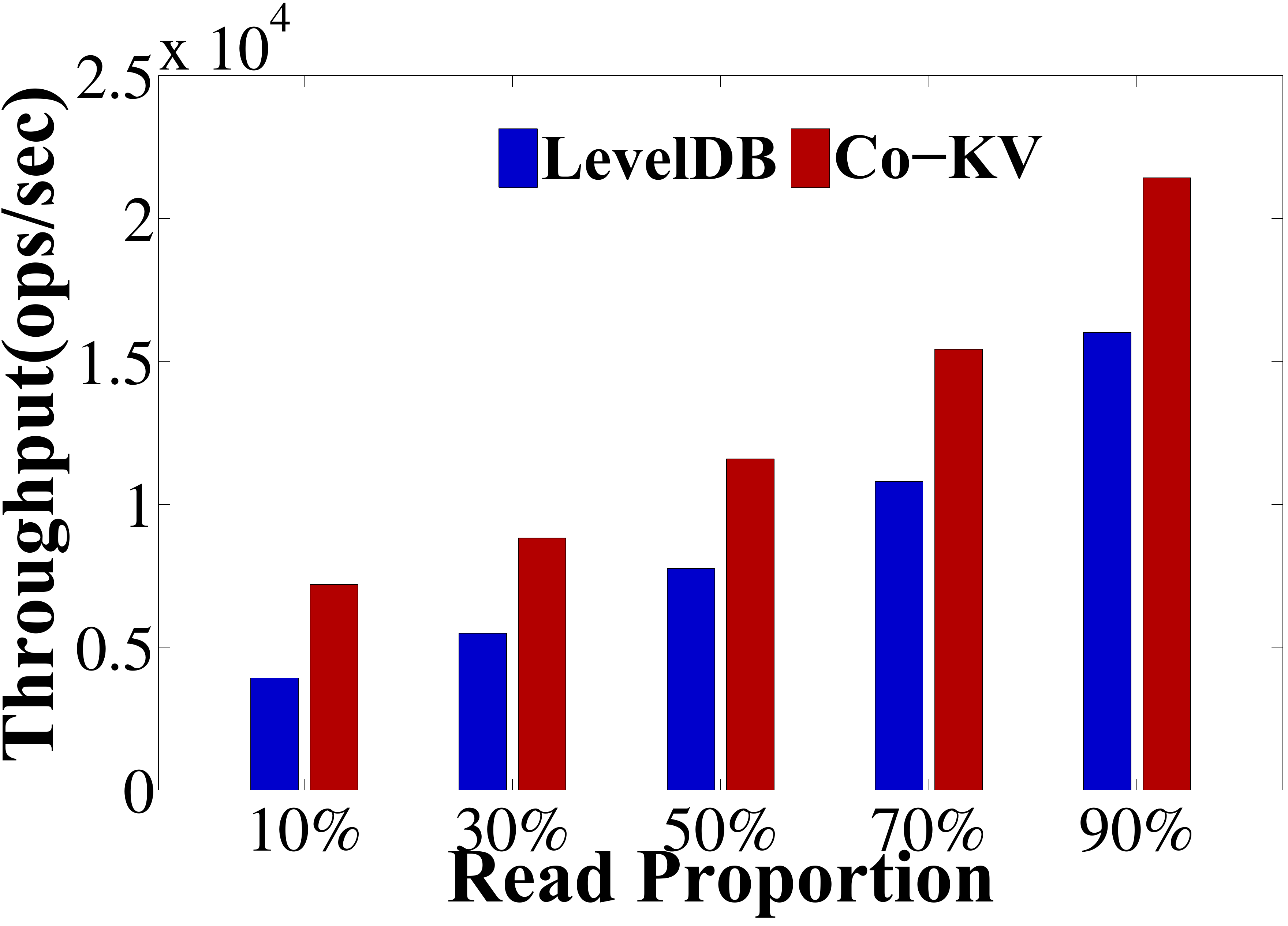}
\label{updatepercent_ops-y3}
}
\caption{Throughput (\ref{OPS-y1}), Write amplification (\ref{WAR-y1}), and Average Latency (\ref{averageLatency-y1}) for LevelDB and Co-KV under YCSB workloads (load 20GB and run 20GB, 40GB, 60GB, 80GB, and 100GB). Throughput (\ref{updatepercent_ops-y3}) for LevelDB and Co-KV based on different read and update ratio under YCSB.}
\vspace{-18pt}
\end{figure*}
In this section, we compare our Co-KV with LevelDB to analyze the experimental results in the aspects of write amplification and throughput under db\_bench workload. 

%%%%%%%%%%%%%%
\textit{{\textbf{Write Amplification}}}: Performance of a key-value store is largely dependent on write amplification, which is defined as the average of actual amount of written data to disk per user written data. Its ideal value is one. There are two kinds of compaction tasks in the \textit{Co-K}$V_H$ and the \textit{Co-K}$V_D$, respectively. Write amplification, a terminology in the user level, is considered in \textit{Co-K}$V_H$ rather than \textit{Co-K}$V_D$. Write amplification is much smaller in Co-KV than that in LevelDB. 
%In this experiment, we use five kinds of data sets in db\_bench\_1 ({\em i.e.,} 200 million, 400 million, 600 million, 800 million, and a billion key-value items) under two setting (gcc -g2 and gcc -o2). During a compaction, an SSTable from the low level ({\em {\em e.g.,}} $L_{k}$) should be pushed down to the higher one ({\em e.g.,} $L_{k+1}$). From our experiment in LevelDB, more than 14 SSTables in the worst case are involved in the compaction. For example, there are 15 new SSTables after the merge-sort operation for these SSTables. These new SSTables are rewritten to the high level, which means that the write amplification is 15 under this condition. Write amplification becomes much higher, ($L*15$) for example, along with increase of data volume and SSTable level. When the level number reaches to 7, the worst value of write amplification is up to 105. Then, much CPU and I/O resources are largely consumption in a key-value store, which can efficiently serve a user request.
In this experiment, we use five kinds of data sets in db\_bench\_1 ({\em i.e.,} 200 million, 400 million, 600 million, 800 million, and a billion key-value items). 
In Figure \ref{seq-random-WAR}, write amplification in LevelDB rises with growing scale of the data set under fillrandom workload in db\_bench\_1. For a size of data set of 1 billion, the write amplification for LevelDB reaches up to 20.7. In contrast, the Co-KV store improves the value by more than 36.0\%. In Co-KV, compaction tasks are offloaded to the \textit{Co-K}$V_D$ manager in the device by means of a compaction-SSTables-aware scheme. Compaction in the \textit{Co-K}$V_H$ manager and the \textit{Co-K}$V_D$ manager is performed in parallel, and compaction in device will not increase host-level write amplification. In db\_bench\_1, \textit{Co-K}$V_D$ significantly reduces write amplification of the overall key-value store. 

It is required/necessary that compaction does not occur under sequential fillseq workload (see db\_bench\_2 in Table II). There is no need to offload compaction to \textit{Co-K}$V_D$. Write amplification in Co-KV is depended on \textit{Co-K}$V_H$ similar to that in LevelDB. Therefore, write amplification is of the same value ({\em i.e.,} 2.1) in Co-KV and in LevelDB under this workload (see Figure \ref{seq-random-WAR}). The value is higher than the ideal reference ({\em i.e.,} one) because of meta-data update. 
%%%%%%%%%%%%%%%
%\begin{figure}[!t]
%\centering
%\includegraphics[height=1.7in, width=2.3in]{fillseqWAR100gb}
%\caption{Write Amplification Ratio under the fillseq. (gcc -g2)}
%\label{Seq-WA}
%\end{figure}
%%%%%%%%%%%%%%%%%%%%%%%%%%%%%%%%%%%
%\begin{figure}[!t]
%\centering
%\includegraphics[height=1.7in, width=2.3in]{fillseqWAR100gb}
%\caption{Write Amplification Ratio under the fillseq workload}
%\label{Seq-WA}
%\end{figure}
%%%%%%%%%%%%%%%%%%%%

\textit{\textbf{Throughput (MB/s)}}: Throughput of key-value store is largely determined by write amplification. Write amplification decreases considerably and throughput is improved as well.   
We employ the same workloads in db\_bench (see Table II) in this test.	
Db\_bench\_1 and db\_bench\_2 have identical data sets from 200 million to one billion key-value items with a key size of 16Bytes and a value size of 100Bytes.
%random -g2
%\begin{figure}[!t]
%\centering
%\includegraphics[height=2.1in, width=2.6in]{fillseq-g2MB}
%\caption{Throughput under the fillseq (gcc -g2).}
%\label{fillseqgMB}
%\end{figure}
%random -o2
%\begin{figure}[!t]
%\centering
%\includegraphics[height=1.8in, width=2.4in]{fillseqOMB}
%\caption{Throughput under the fillseq workload.}
%\label{fillseqOMB}
%\end{figure}
%\begin{figure}[!t]\footnotesize
%\centering
%\subfigure[]{
%\includegraphics[height=0.7in, width=0.90in]{fillseqOMB}
%\label{fillseqOMB}
%}
%\subfigure[]{
%\includegraphics[height=0.7in, width=0.90in]{fillrandomOMB}
%\label{fillrandoMB}
%}
%\caption{Throughput (MB/s) for LevelDB and Co-KV under fillseq and fillrandom workloads, \ref{fillseqOMB} fillseq workload and \ref{fillrandoMB} fillrandom workload.}
%\vspace{-15pt}
%\end{figure}
Results under fillseq workload in db\_bench\_2 are shown in Figure \ref{fillseqOMB}. The maximum throughputs for LevelDB and Co-KV rise up to 48.7MB/s and 48.9MB/s, respectively. 
%Key ranges in SSTables are no-overlapping; and then there is no compaction. Performance for Co-KV is largely dependent on \textit{Co-K}$V_H$, which is similar to LevelDB. Throughput for Co-KV and LevelDB has similar values under this workload. 
Under fillseq workload, compaction will not be triggered in Co-KV and LevelDB; thus, they have similar throughputs.
Figure \ref{fillrandoMB} presents the throughputs of Co-KV and LevelDB under fillrandom workload. The size of the data set is approximately between 20GB and 110GB. Throughput for LevelDB ranges from 2.0MB/s to 3.1MB/s.

In contrast, the throughput in Co-KV is about twice as great as that in LevelDB. The value ranges from 3.5MB/s to 5.9MB/s, with a best improvement of about 2.0x.
As data volume increases, much more compaction is triggered and write amplification becomes higher. In this case, throughput is degraded both in Co-KV and in LevelDB. Co-KV reduces write amplification by up to 36.0\% due largely to its parallel and collaborative compactions in \textit{Co-K}$V_H$ and \textit{Co-K}$V_D$. It is concluded that Co-KV relieves the run-time cost in the process of compaction to achieve better throughput compared with LevelDB.
%%seq -g2
%\begin{figure}[!t]
%\centering
%\includegraphics[height=1.8in, width=2.4in]{fillrandom-g2MB}
%\caption{Throughput under the fillrandom (gcc -g2).}
%\label{fillrandgMB}
%\end{figure}
%seq -o2
%\begin{figure}[!t]
%\centering
%\includegraphics[height=2in, width=2.4in]{fillrandomOMB}
%\caption{Throughput under the fillrandom workload.}
%\label{fillrandoMB}
%\end{figure}

%我们使用上述一样的实验配置，观察LevelDB和COL-KV在不同负载下（fillseq,fillrandom）的cpu使用率。其中，cpu 使用率只统计了主机端进程的cpu使用率。  
%如图xx，可以看出LevelDB在fillseq的负载下，和COL-KV的cpu使用率几乎相同。该负载由于没有引起compaction，COL-KV和LevelDB使用了相同的处理机制。然而，在fillrandom下，COL-KV将部分的计算迁移到device端，因此，基于xx KV items 情况下，COL-KV的主机端cpu使用率分别为，xxxxxxxx 较LevelDB更小。在数据集为1billion的KV项时，LevelDB和COL-KV的cpu使用率分别为13.8%,10.0%；

\subsubsection{{\textbf{Experiment based on YCSB}}}

A number of workloads are configured on the basis of varied ratios of read to write, size of data set (under load or run operations), and record size in YCSB. Throughput (operations per second, ops/sec), write amplification, and average latency in YCSB are employed to evaluate Co-KV under YCSB workloads.

\textbf{\textit{Different Run Data Volumes (Load 20GB database, Record size = 1KB)}}: 
%\begin{figure}[!t]
%\centering
%\includegraphics[height=1.5in, width=2.8in]{OPS}
%\caption{Throughput for LevelDB and Co-KV under YCSB}
%\label{ycsb-ops}1.0in, width=1.3in
%\end{figure}
A workload (in Table II) with 100 percent update operations and Zipf distribution is configured in YCSB. The record size is 1KB, load 20GB database and five types of databases are configured during running. 
As shown in Figure \ref{OPS-y1}, Co-KV improves the throughput (ops/sec) to 2.0x at the least (run 100GB database) and 2.2x at the most (run 20GB database) compared with LevelDB. 100\% update operations in this workload can cause lots of compaction for LevelDB or Co-KV to degrade its throughput. In Co-KV, the device-side \textit{Co-K}$V_D$ runs about half of the compaction operations in LevelDB. The host- and device-side compaction operations can be executed in parallel. These features of Co-KV contribute to its better throughput than that in LevelDB.

Similarly, write amplification is noticeably improved in Co-KV in comparison with that in LevelDB (see Figure \ref{WAR-y1}). In the best case, write amplification in Co-KV is decreased to about 71.0\% what it is in LevelDB under load 20GB and run 80GB database based workload. Note that write amplification degrades the overall performance. As Figure \ref{averageLatency-y1} shows, the average latency of update in Co-KV is optimized by about 50.0\% $\sim$ 54.0\% compared to that in LevelDB.

\textbf{\textit{Different Read and Update Ratios (Load 10GB, Run 20GB database, Record size = 1KB)}}: 		
%height=1.5in, width=2.8in
%\begin{figure}[!t]
%\setlength{\abovecaptionskip}{0.cm}
%\setlength{\belowcaptionskip}{-0.cm}
%\centering
%\includegraphics[height=0.92in, width=1.3in]{updatepercent_ops-y3}
%\caption{Throughput for LevelDB and Co-KV based on different read and update ratio under YCSB.}
%\label{updatepercent_ops-y3}
%\vspace{-18pt}
%\end{figure}
In the experiment, we configure the ratio of read and update operations (get/put) in a workload to evaluate the throughput for Co-KV. This workload has 10GB load database, 20GB of run database, and 1KB of record size. The read operation is configured as 10\%, 30\%, 50\%, 70\%, and 90\%, respectively. The throughput in Co-KV shows to be 7186ops/s, 8820ops/s, 11578ops/s, 15428ops/s, and 21415ops/s under 10\%-, 30\%-, 50\%-, 70\%-, and 90\%-read workloads. Although the throughput in LevelDB has the same tendency, the value of throughput (ops/sec) is smaller than that in Co-KV under the same workload. Firstly, the update operations in Co-KV or LevelDB induce compaction which affects its throughput. More update operations (such as 90\% update) can trigger compaction which cuts down the throughput for both Co-KV and LevelDB. Secondly, Co-KV with improved compaction has better throughput than LevelDB (see Figure \ref{updatepercent_ops-y3}).

\section{Discussion and Future Work} 
\label{discussion}
We hope that this work can provide insights into a new research field of compaction improvement using collaboration of host and device for LSM-tree based key-value stores in perspective of system-level parallelism. Our Co-KV is also applied into other LSM-tree based key-value stores on an SSD capable of near-data processing. Characterized by a compaction offloading scheme and the semantic management between host and device, Co-KV makes transparent to users the subsystem with the NDP model in the device, and is fully compatible with the application interfaces in LSM-tree based storage systems. Therefore, Co-KV enjoys flexibility for key-value storage system. 

In this work, the \textit{Co-K}$V_H$ is derived from LevelDB and demonstrates a relatively large compaction improvement in the host side. One of our future work is to design and implement a new dynamic \textit{Co-K}$V_H$-based scheme to improve compaction in the host for Co-KV. In addition, we will primarily address the implementation complexity in the Offload manager, \textit{Co-K}$V_H$ manager, \textit{Co-K}$V_D$ manager, and semantic manager to deploy Co-KV on a real-world platform in the future. 

\section{Conclusion}
\label{conclusion}
Compaction has an adverse effect on the performance of LSM-tree based key-value stores. Existing work mostly concentrates on alleviating compaction in the host level. The collaborative improvement for compaction is rarely taken into account. We propose a system-level Co-KV to minimize the consumption of compaction in the LSM-tree based key-value stores. It employs the near-data processing model in the device and takes full advantage of the collaborative computation on both host and device. Co-KV divides a compaction into two portions, which are offloaded into host and device, respectively. Collaborative compactions are performed in parallel. Our Co-KV achieves better performance than LevelDB in terms of write amplification, throughput, and latency under db\_bench and YCSB workloads.
\section*{Acknowledgment}
{We thank the reviewers for their work on our paper. 
This work is supported in part by National Natural Science Foundation of China under Grants 61702004.}
\bibliographystyle{IEEEtran}\scriptsize
% argument is your BibTeX string definitions and bibliography database(s)
%\bibliography{sh}

% that's all folks
\end{document}